\documentclass[11pt]{article}

\usepackage{latexsym}
\usepackage{amsfonts}
\newtheorem{theorem}{Theorem}
\newtheorem{lemma}[theorem]{Lemma}
\newtheorem{corollary}[theorem]{Corollary}
\newcommand{\binom}[2]{\left({{#1}\atop{#2}}\right)}

\newenvironment{proof}{\trivlist\item[\hskip \labelsep{\textit{Proof.}}]}%
                 {\hfil\hfill$\square$\endtrivlist}

\begin{document}

\title{Unique characterization of the Bel-Robinson tensor}

\author{G Bergqvist$^1$ and P Lankinen$^2$ \\
${}^1$Matematiska institutionen, Link\"opings universitet, \\
      SE-581 83 Link\"oping, Sweden \\
${}^2$Department of Mathematics and Physics, M\"alardalen University, \\
SE-721 23 V\"aster{\aa}s, Sweden \\
gober@mai.liu.se, paul.lankinen@mdh.se}

\maketitle

\begin{abstract}
We prove that a completely symmetric and trace-free rank-4 tensor is, up to sign, 
a Bel-Robinson type tensor, i.e., the superenergy tensor
of a tensor with the same algebraic symmetries as the Weyl tensor, if and only if it 
satisfies a certain quadratic identity. This may be seen as the first Rainich theory 
result for rank-4 tensors.
\end{abstract}

%\pacs{04.20.Cv, 04.20.Gz, 02.40.Ky}

\newpage

\section{Introduction}

In classical Rainich(-Misner-Wheeler) theory the following is proven assuming dimension four
and Lorentzian metric \cite{BH,BS,MW,R}:

\smallskip

\begin{theorem}\label{th:RaiEM}
A symmetric trace-free tensor $T_{ab}$ which satisfies the dominant energy condition
can be written $T_{ab}=-{1\over 2}(F_{ac}F_{b}{}^c+{}^*F_{ac}{}^*F_{b}{}^c)\equiv -F_{ac}
F_{b}{}^c+{1\over 4} g_{ab}F_{cd}F^{cd}$, where $F_{ab}$
is a 2-form, if and only if 
$$
T_{ac}T_{b}{}^c={1\over 4}g_{ab}T_{cd}T^{cd}\ .
$$
\end{theorem}

\smallskip
Recall that the dominant energy condition is $T_{ab}u^a v^b\ge 0$ for all future-directed
causal vectors $u^a$ and $v^a$. By ${}^*F_{ab}$ we mean the dual 2-form of $F_{ab}$.

The theorem means that $T_{ab}$ is algebraically the energy-momentum tensor of a Maxwell
field $F_{ab}$. One may also, using Einstein's equation, replace $T_{ab}$ by $R_{ab}$ 
in the statement. The result may then be interpreted as giving necessary and suffcient
conditions on a geometry to correspond to an Einstein-Maxwell spacetime physically.

There have been various generalizations of this result. In \cite{BS} it was shown in arbitrary
dimension that a symmetric tensor $T_{ab}$ which satisfies the dominant energy condition
can be written as the superenergy tensor of a simple $p$-form \cite{S}, 
$T_{ab}={{(-1)^{(p-1)}}\over{(p-1)!}}(\Omega_{ac\dots d}\Omega_{b}{}^{c\dots d}-{1\over {2p}}g_{ab}\Omega_{ec\dots d}\Omega^{ec\dots d})$, where $\Omega_{ac\dots d}$
is a simple $p$-form, if and only if $T_{ac}T_{b}{}^c={1\over 4}g_{ab}T_{cd}T^{cd}$.
That a $p$-form is simple means that it is a wedge product of $p$ 1-forms. Furthermore,
the trace of $T_{ab}$ determines $p$. Some special cases of this result were already
known. It was also shown that the dominant energy condition could be removed since
$T_{ac}T_{b}{}^c={1\over 4}g_{ab}T_{cd}T^{cd}$ implies that either $T_{ab}$ or $-T_{ab}$
satisfies the dominant energy condition. Therefore the conclusion without the dominant
energy condition is 
$\pm T_{ab}={{(-1)^{(p-1)}}\over{(p-1)!}}(\Omega_{ac\dots d}\Omega_{b}{}^{c\dots d}-{1\over {2p}}g_{ab}\Omega_{ec\dots d}\Omega^{ec\dots d})$.

In \cite{BH} superenergy tensors of more 
general p-forms were considered and the results 
of \cite{BS} were generalized in the way that the condition 
$T_{ac}T_{b}{}^c={1\over 4}g_{ab}T_{cd}T^{cd}$
was replaced by a third-order equation for $T_{ab}$.

The classical result has a very natural formulation in terms of spinors. That
$T_{ab}=-F_{ac}F_{b}{}^c+{1\over 4}g_{ab}F_{cd}F^{cd}$ can in terms of spinors
be written $T_{ab}=2\varphi_{AB}\bar\varphi_{A'B'}$ where $\varphi_{AB}
=\varphi_{(AB)}$ is a symmetric spinor which represents the Maxwell field. 
In fact a purely spinorial proof is
the simplest way to demonstrate the classical result. In \cite{BH} and \cite{BS} tensorial
methods were used to find the generalizations.

Until now no Rainich type results have been presented for higher rank superenergy tensors
\cite{S} but the aim here is to prove a first such result. The result is for the most well-known of all
rank-4 superenergy tensors, the Bel-Robinson tensor, and is the following:

\smallskip
\begin{theorem}\label{th:RaiBR}
In four dimensions, a completely symmetric and trace-free rank-4 tensor 
$T_{abcd}$  is, up to sign, a Bel-Robinson type tensor,
i.e. $\pm T_{abcd}=C_{akcl}C_b{}^k{}_d{}^l+{}^*C_{akcl}{}^*C_b{}^k{}_d{}^l$ 
where $C_{abcd}$ has the same algebraic symmetries as the Weyl tensor, if and only if
\begin{eqnarray}\label{eq:Rai}
T_{jabc}T^{jefg}=&{3\over 2}g_{(a}{}^{(e}T_{bc)jk}T^{fg)jk}+
{3\over 4}g_{(a}{}^{(e}T_{|jk|b}{}^{f}T_{c)}{}^{g)jk}
-{3\over 4}g_{(ab}T_{c)jk}{}^{(e}T^{fg)jk}
\nonumber \\
-&{3\over 4}g^{(ef}T_{jk(ab}T_{c)}{}^{g)jk}
+{1\over 32}(3g_{(ab}g_{c)}{}^{(e}g^{fg)}-4g_{(a}{}^{(e}g_b{}^f g_{c)}{}^{g)})T_{jklm}T^{jklm}
\end{eqnarray}
\end{theorem}

\smallskip

This may also be stated as $T_{abcd}$ is the superenergy tensor of a Weyl candidate 
tensor  (which means a tensor with same algebraic symmetries as the Weyl tensor:
$C_{abcd}=-C_{bacd}=-C_{abdc}=C_{cdab},\ C_{abcd}+C_{adbc}+C_{adcb}=0,\ 
C^a{}_{bad}=0$). The theorem is a natural generalization of the classical Rainich 
theory as the Bel-Robinson tensor in terms of spinors can be written 
$$
T_{abcd}=4\Psi_{ABCD}\bar\Psi_{A'B'C'D'}
$$
where $\Psi_{ABCD}=\Psi_{(ABCD)}$ is the Weyl spinor. In the proof we shall see that
the condition (\ref{eq:Rai}) in Theorem \ref{th:RaiBR} equivalently can be replaced by
\begin{eqnarray}\label{eq:RaiS}
T_{jbc(a}T_{e)}{}^{jfg}=&g_{(b}{}^{(f}T_{c)jk(a}T_{e)}{}^{g)jk}-
{1\over 4}g^{fg}T_{jkb(a}T_{e)c}{}^{jk}-
{1\over 4}g_{bc}T_{jk}{}^{f}{}_{(a}T_{e)}{}^{gjk}
\nonumber \\
+&{1\over 4}g_{ae}(T_{jkbc}T^{jkfg}+{1\over 8}(g_{bc}g^{fg}-g_b{}^f g_c{}^g-
g_b{}^g g_c{}^f )T_{jklm}T^{jklm})
\end{eqnarray}
This is the symmetric part of (\ref{eq:Rai}) with respect to $ae$, hence the anti-symmetric
part gives no additional information but (\ref{eq:Rai}) might be considered a more
natural identity than (\ref{eq:RaiS}) from the point of view of index symmetries.

Note that taking a trace of (\ref{eq:Rai}) or (\ref{eq:RaiS}) one finds as a necessary (but not
sufficient) condition
\begin{equation}\label{eq:BR3c}
T_{jkla}T_e{}^{jkl}={1\over 4}g_{ae}T_{jklm}T^{jklm}
\end{equation}
which is a well-known
identity for the Bel-Robinson tensor \cite{PR}. Here we especially remark that this is obtained
by taking only one trace of (\ref{eq:Rai}) as all terms with two contractions (of type $T_{jkab}T^{jkef}$) 
then cancel and only some with three or four contractions remain. Thus no further identity, which would 
have been necessary but not sufficient, 
between (\ref{eq:Rai}) and  (\ref{eq:BR3c}) exists for the Bel-Robinson tensor
(although equation (\ref{eq:as4}) below is another type of necessary but not sufficient identity).

By the {\it dominant property} we mean the following generalization of the dominant
energy condition: 
\begin{equation}\label{dp}
T_{abcd}u^av^bw^cz^d\ge 0
\end{equation}
 for all causal  future-directed vectors $u^a$, $v^a$, $w^a$ and $z^a$, and any tensor
 having this property is called a {\it causal} tensor. Since the Bel-Robinson
 tensor has the dominant property \cite{B} we get the following
 \smallskip
\begin{corollary}\label{cor1}
If $T_{abcd}$  is completely symmetric, trace-free, and satisfies (\ref{eq:Rai})
then either $T_{abcd}$ or $-T_{abcd}$ has the dominant property.
\end{corollary}
\smallskip
If the dominant property is added explicitly as a condition, then clearly the + sign is choosen
in Theorem \ref{th:RaiBR}
and we can, in a way similar to Theorem  \ref{th:RaiEM}, formulate
\smallskip
\begin{corollary}\label{cor 2}
A completely symmetric and trace-free rank-4 tensor 
$T_{abcd}$ which satisfies the dominant property is a Bel-Robinson type tensor,
i.e. $T_{abcd}=C_{akcl}C_b{}^k{}_d{}^l+{}^*C_{akcl}{}^*C_b{}^k{}_d{}^l$ 
where $C_{abcd}$ has the same algebraic symmetries as the Weyl tensor, if and only if
(\ref{eq:Rai}) is satisfied.
\end{corollary}
\smallskip
Another result which follows immediately from (\ref{eq:Rai}) and is non-trivial to prove by 
other tensor methods is 
\smallskip
\begin{corollary}\label{cor 2.5}
$T_{abcd}\ell^a\ell^b n^c$ is null whenever $\ell^a$ and $n^a$ are null.
\end{corollary}
\smallskip
To see this just contract (\ref{eq:Rai}) with $\ell^a\ell^b n^c\ell_e\ell_f n_g$ and the non-vanishing 
terms on the right-hand side trivially cancel out. Even easier follows the special case that
$T_{abcd}\ell^a\ell^b\ell^c$ is null whenever $\ell^a$ is null since 
contracting (\ref{eq:Rai}) with $\ell^a\ell^b\ell^c\ell_e\ell_f\ell_g$,
each term on the right-hand side vanishes.

Hence we have got a very simple tensorial proof of Corollary \ref{cor 2.5}. Less direct is to 
prove that $T_{abcd}\ell^a n^b k^c$ is null if $\ell^a$, $n^a$ and $k^a$ are null 
(note that this leads to a proof of the dominant property \cite{B}).

Our methods of proving the theorem will be spinorial, thus extending the simplest
way of proving the classical rank-2 case. It seems that the tensorial methods
used in \cite{BH} and \cite{BS} are very complicated to generalize to the higher rank case.

Theorem \ref{th:RaiEM} and the generalizations presented above are usually called
{\it algebraic} Rainich (type) conditions \cite{PR}. This simply refers to that the tensors
satisfy polynomial relations. For the tensors also to correspond to a field satisfying the
Maxwell or some other field equation one can derive so-called {\it differential} Rainich
(type) conditions  \cite{PR}. For the Bel-Robinson tensor and other higher rank
superenergy tensors such results will be presented in future work.

Our result in Theorem \ref{th:RaiBR} represents a fundamental property of the Bel-Robinson
tensor, a tensor which nowadays is maybe the most important quantity in the study of
the Cauchy problem for Einstein's vacuum equations. The search for the identity 
(\ref{eq:Rai}) has been proposed by various people. It has not been known
either whether such an identity would also be sufficient in the sense we prove in 
Theorem \ref{th:RaiBR}, or if there
would be further identities from traces of (\ref{eq:Rai}) as we show there are not besides
the already known (\ref{eq:BR3c}). We also
see our result as the first in a more general study of relations between higher rank
superenergy tensors and causal tensors, in a way similar to the rank-2 case
developed in \cite{BS} in which the corresponding identity 
$4T_{aj}T_{b}{}^j=g_{ab}T_{jk}T^{jk}$ plays a fundamental role.

In section 2 we review some basic results about 2-spinors, especially concerning
symmetrization and antisymmetrization techniques. To illustrate the methods
we will use to prove Theorem \ref{th:RaiBR} we also present the proof of the rank-2 case on a 
form suitable for generalizations to the much more complex rank-4 case.
In section 3 we then prove the theorem for the Bel-Robinson tensor.

\section{Basic spinor properties and the rank-2 case}

\subsection{Basic properties of 2-spinors}

We recall here some well-known facts about spinors, especially related to symmetrization
and antisymmetrization. The formulas can be found in the book by Penrose and Rindler
\cite{PR} and we also follow their notation and conventions (except for a factor 4 in the definition
of the  Bel-Robinson tensor). Spinor expressions for general superenergy tensors are
given in \cite{B}.

We use $A,B,\dots ,A',B',\dots$ for spinor indices and identify with tensor indices
$a,b,\dots$ according to $AA'=a$. A spinor $P_{AB\mathcal{Q}}$ , where $\mathcal{Q}$ represents some 
set of spinor indices, can be divided up into its 
symmetric and antisymmetric parts with respect to a pair of indices
$$
P_{AB\mathcal{Q}}={1\over 2}(P_{AB\mathcal{Q}}+P_{BA\mathcal{Q}})+
{1\over 2}(P_{AB\mathcal{Q}}-P_{BA\mathcal{Q}})=P_{(AB)\mathcal{Q}}+P_{[AB]\mathcal{Q}} \ .
$$
The antisymmetric part can be written
$$
P_{[AB]\mathcal{Q}}={1\over 2}\varepsilon_{AB}P_{C}{}^C{}_{\mathcal{Q}} \ ,
$$
where $\varepsilon_{AB}=-\varepsilon_{BA}$, so
\begin{equation}\label{eq:sas}
P_{AB\mathcal{Q}}=P_{(AB)\mathcal{Q}}+{1\over 2}\varepsilon_{AB}P_{C}{}^C{}_{\mathcal{Q}}\ .
\end{equation}
From this one also has
\begin{equation}\label{eq:sas2}
P_{AB\mathcal{Q}}=P_{BA\mathcal{Q}}+\varepsilon_{AB}P_{C}{}^C{}_{\mathcal{Q}} \ .
\end{equation}
A simple but very useful rule is
\begin{equation}\label{eq:zz}
P_{C}{}^C{}_{\mathcal{Q}}=-P^C{}_{C\mathcal{Q}} \ .
\end{equation}
Note that if $P_{ab\mathcal{Q}}=P_{ba\mathcal{Q}}$ then we have
$$
P_{BAA'B'\mathcal{Q}}=P_{ab\mathcal{Q}}-{1\over 2}g_{ab}P_{c}{}^{c}{}_{\mathcal{Q}} \ ,
$$
where $g_{ab}=\varepsilon_{AB}\bar\varepsilon_{A'B'}$ so permuting $A$ and $B$ gives a trace reversal. From this we find another formula 
we shall need
\begin{equation}\label{eq:ds}
P_{(AB)(A'B')\mathcal{Q}}=P_{(ab)\mathcal{Q}}-{1\over 4}g_{ab}P_{c}{}^{c}{}_{\mathcal{Q}} \ .
\end{equation}
The relation between a 2-form $F_{ab}$ and a symmetric spinor $\varphi_{AB}$ is
$$
F_{ab}=\varphi_{AB}\bar\varepsilon_{A'B'}+\bar\varphi_{A'B'}\varepsilon_{AB} 
\qquad ; \qquad \varphi_{AB}={1\over 2}F_{AC'B}{}^{C'} 
$$
and one also has
$$
-F_{ac}F_{b}{}^c+{1\over 4}g_{ab}F_{cd}F^{cd}=2\varphi_{AB}\bar\varphi_{A'B'} \ . 
$$
For the Weyl tensor $C_{abcd}$ and the completely symmetric Weyl spinor
$\Psi_{ABCD}$ the corresponding relations are
\begin{equation}\label{eq:w1}
C_{abcd}=\Psi_{ABCD}\bar\varepsilon_{A'B'}\bar\varepsilon_{C'D'}+
\bar\Psi_{A'B'C'D'}\varepsilon_{AB}\varepsilon_{CD}
\qquad ; \qquad \Psi_{ABCD}={1\over 4}C_{AE'B}{}^{E'}{}_{CF'D}{}^{F'}  \ .
\end{equation}
and
\begin{equation}\label{eq:w2}
C_{akcl}C_b{}^k{}_d{}^l+{}^*C_{akcl}{}^*C_b{}^k{}_d{}^l=4\Psi_{ABCD}\bar\Psi_{A'B'C'D'} \ . 
\end{equation}

\smallskip

We will study completely symmetric and trace-free tensors $T_{a\dots b}$ . These
two properties together are very elegantly expressed in an equivalent way using 
spinor indices as
$$
T_{a\dots b}=T_{(A\dots B)(A'\dots B')} \ .
$$
If  a tensor $T_{a\dots b}$ can be written
\begin{equation}\label{eq:fa}
T_{a\dots b}=\chi_{A\dots B}\bar\chi_{A'\dots B'} \ ,
\end{equation}
for some completely symmetric spinor $\chi_{A\dots B}=\chi_{(A\dots B)}$, then it 
follows trivially that $T_{a\dots b}$ is (i) completely symmetric, (ii) trace-free, (iii)
satisfies the dominant property (\ref{dp}), and (iv)
 \begin{equation}\label{eq:fun}
T_{A\dots B}^{A'\dots B'}T_{C\dots D}^{C'\dots D'}=
 T_{A\dots B}^{C'\dots D'}T_{C\dots D}^{A'\dots B'} \ .
\end{equation}
 Conversely, suppose that $T_{a\dots b}$ has properties (i), (ii) and (iv). Let $u^a,\dots , v^a$
 be future-directed null vectors such that $T_{a\dots b}u^a\dots v^b=k\ne 0$. Such
 null vectors must exist since otherwise, by taking linear combinations,  we would get 
 $T_{a\dots b}u^a\dots v^b=0$ for all 
 vectors which would imply $T_{a\dots b}=0$ . Then write the null vectors in terms
 of spinors as $u^a=\alpha^A\bar\alpha^{A'}, \dots ,v^a=\beta^A\bar\beta^{A'}$.
 Contract  (\ref{eq:fun}) with these spinors to get
 $$
T_{A\dots BA'\dots B'}T_{C\dots DC'\dots D'}\alpha^C\bar\alpha^{C'}\dots
  \beta^D\bar\beta^{D'}=
 (T_{A\dots BC'\dots D'}\bar\alpha^{C'}\dots\bar\beta^{D'})(
 T_{C\dots DA'\dots B'}\alpha^{C}\dots\beta^{D}) 
$$
 from which follows that either $T_{a\dots b}$ or $-T_{a\dots b}$ can be 
 factorized as in (\ref{eq:fa}) with $\chi_{A\dots B}=
 {1\over \sqrt{|k|}}T_{A\dots BC'\dots D'}\bar\alpha^{C'}\dots\bar\beta^{D'}$.
 Hence also (iii) is satisfied for $T_{a\dots b}$ or $-T_{a\dots b}$ and 
 a completely symmetric  and trace-free tensor 
 can, up to sign, be factorized according to (\ref{eq:fa}) if and only if (\ref{eq:fun}) is satisfied.
 
 In this paper we only study symmetric and trace-free tensors but
 note that, more generally, from the above it is also clear that (\ref{eq:fa}) and  
 (\ref{eq:fun}) are equivalent, up to sign in (\ref{eq:fa}), even if no symmetry or 
 trace properties of  $T_{a\dots b}$ are assumed.

\subsection{The spinorial proof of the rank-2 case}

We now use the techniques of Section 2.1 to prove Theorem \ref{th:RaiEM}. 
We do it without assuming the dominant energy condition so the conclusion will
be $\pm T_{ab}=-{1\over 2}(F_{ac}F_{b}{}^c+{}^*F_{ac}{}^*F_{b}{}^c)$. We
essentially follow the proof given in \cite{PR} but write it in a way suitable for
generalizations to higher rank. It is clear that what must be proven is that
$T_{ab}=T_{(AB)(A'B')}$  satisfies 
$T_{ac}T_{b}{}^c={1\over 4}g_{ab}T_{cd}T^{cd}$
if and only if $\pm T_{ab}=\varphi_{AB}\bar\varphi_{A'B'}$ for a symmetric $\varphi_{AB}$.
By the above argument, this factorization is now equivalent to 
 \begin{equation}\label{eq:fun2}
T_{AB}^{A'B'}T_{CD}^{C'D'}- T_{AB}^{C'D'}T_{CD}^{A'B'}=0 \ .
\end{equation}
 To study this equation, we begin by dividing up the left-hand side into symmetric
 and anti-symmetric parts with respect to the pairs $A'D'$ and $B'C'$. The antisymmeric
 parts give contractions so we get three types of terms: with two symmetrizations and
 no contraction, with one symmetrization and one contraction, and with no symmetrization
 and two contractions. The first type looks like
$$
T_{AB}^{(A'|(B'}T_{CD}^{C')|D')}- T_{AB}^{(C'|(D'}T_{CD}^{A')|B')}
$$
which obviously vanishes. (Here we use the standard notation $(A|\dots |B)$ to denote 
symmetrization over $AB$ but not over indices written between $A$ and $B$.)
The second type is (without the $\varepsilon^{A'D'}$ written out)
$$
T_{ABK'}^{(B'}T_{CD}^{C')K'}- T_{AB}^{K'(C'}T_{CDK'}^{B')}
$$
which by (\ref{eq:zz}) is equal to $2T_{ABK'}^{(B'}T_{CD}^{C')K'}$. The third type is
$$
T_{ABK'L'}T_{CD}^{K'L'}- T_{AB}^{K'L'}T_{CDK'L'} 
$$
 which, by applying (\ref{eq:zz}) twice, vanishes. Therefore  (\ref{eq:fun2}) is equivalent
 to
 \begin{equation}\label{eq:fun21}
T_{ABK'}^{(B'}T_{CD}^{C')K'}=0 \ .
\end{equation}
Taking symmetric and antisymmetric parts of (\ref{eq:fun21}) with respect to the pairs $AD$ 
and $BC$ gives again three types of terms. Symmetrization twice gives
$$
T_{K'(A|(B}^{(B'}T_{C)|D)}^{C')K'}
$$
which vanishes by applying (\ref{eq:zz}). Antisymmetrization (contraction) twice
gives 
$$
T_{KLK'}^{(B'}T_{}^{C')KLK'}
$$
which vanishes by applying (\ref{eq:zz}) three times. Left are terms with one symmetrization
and one contraction. Hence (\ref{eq:fun2}) is equivalent  to
$$
T_{KK'(B}^{(B'}T_{C)}^{C')KK'}=0 \ .
$$
Lowering $B'$ and $C'$ and using (\ref{eq:ds}), this is equivalent to
$$
T_{k(b}T_{c)}{}^{k}={1\over 4}g_{bc}T_{kl}T^{kl} \ .
$$
Since $T_{k(b}T_{c)}{}^{k}=
T_{kb}T_{c}{}^{k}$ we get
$$
T_{kb}T_{c}{}^{k}={1\over 4}g_{bc}T_{kl}T^{kl} 
$$
which completes the proof of Theorem  \ref{th:RaiEM}.

\section{The Bel-Robinson case}

We turn now to the proof of Theorem  \ref{th:RaiBR}. By (\ref{eq:w1}) and (\ref{eq:w2})
Theorem \ref{th:RaiBR} can in terms of spinors equivalently be written
\begin{theorem}\label{th:RaiBR2}
A completely symmetric and trace-free rank-4 tensor 
$T_{abcd}$  can be written $\pm T_{abcd}=\Psi_{ABCD}\bar\Psi_{A'B'C'D'}$ 
with $\Psi_{ABCD}=\Psi_{(ABCD)}$  if and only if
\begin{eqnarray}\label{eq:Rai2}
T_{jabc}T^{jefg}=&{3\over 2}g_{(a}{}^{(e}T_{bc)jk}T^{fg)jk}+
{3\over 4}g_{(a}{}^{(e}T_{|jk|b}{}^{f}T_{c)}{}^{g)jk}
-{3\over 4}g_{(ab}T_{c)jk}{}^{(e}T^{fg)jk}
\nonumber \\
-&{3\over 4}g^{(ef}T_{jk(ab}T_{c)}{}^{g)jk}
+{1\over 32}(3g_{(ab}g_{c)}{}^{(e}g^{fg)}-4g_{(a}{}^{(e}g_b{}^f g_{c)}{}^{g)})T_{jklm}T^{jklm}
 \nonumber
\end{eqnarray}
\end{theorem}
Note that the factor of 4 usually used for the relation between the Bel-Robinson tensor
and the Weyl spinor is irrelevant for the statement and proof of the theorem.

We divide up the proof into some lemmas. 
\begin{lemma}\label{th:L1}
A completely symmetric and trace-free rank-4 tensor 
$T_{abcd}$  can be written $\pm T_{abcd}=\Psi_{ABCD}\bar\Psi_{A'B'C'D'}$ (either + or -)
with $\Psi_{ABCD}=\Psi_{(ABCD)}$  if and only if
\begin{equation}\label{eq:fun4}
T_{ABCD}^{A'B'C'D'}T_{EFGH}^{E'F'G'H'}=T_{ABCD}^{E'F'G'H'}T_{EFGH}^{A'B'C'D'}
\end{equation}
\end{lemma}
\begin{proof}
Obvious from the results in subsection 2.1.
\end{proof}

\begin{lemma}\label{th:L2}
$$
T_{ABCD}^{A'B'C'D'}T_{EFGH}^{E'F'G'H'}=T_{ABCD}^{E'F'G'H'}T_{EFGH}^{A'B'C'D'}
$$
if and only if
$$
T_{J'(D|(C|(B}^{J(D'|(C'|(B'}T_{F)|G)|H)J}^{F')|G')|H')J'}=0 \ , \ 
T_{J'K'L'(D|(C|(B}^{J(B'}T_{F)|G)|H)J}^{F')J'K'L'}=0  \ \ 
  {\mbox{and}} \ \ 
T_{J'K'L'(B}^{JKL(B'}T_{F)JKL}^{F')J'K'L'}=0 
$$
\end{lemma}

\begin{proof}

Let us symmetrize, using (\ref{eq:sas}), the expression 
$$
        T_{ABCD}^{A'B'C'D'}T_{EFGH}^{E'F'G'H'}-T_{ABCD}^{E'F'G'H'}T_{EFGH}^{A'B'C'D'}
$$
with respect to a number of
pairs of indices, upper or lower and contract in the pairs of
indices that are not symmetrized. To start with we disregard the
lower indices and symmetrize in the upper indices. If we
symmetrize in all 4 pairs, there will be no contractions so we get

$$
        T_{ABCD}^{(D'|(C'|(B'|(A'}T_{EFGH}^{E')|F')|G')|H')}-
        T_{ABCD}^{(H'|(G'|(F'|(E'}T_{EFGH}^{A')|B')|C')|D')}=0
$$
due to that we can permute the primed indices pairwise in the second term. Next let us
symmetrize in 3 pairs of indices, then we need to contract in one
pair giving

\begin{equation}\label{type30symmetry}
        T_{J'ABCD}^{(C'|(B'|(A'}T_{EFGH}^{E')|F')|G')J'}-
        T_{ABCD}^{J'(G'|(F'|(E'}T_{EFGHJ'}^{A')|B')|C')}=\
        2T_{J'ABCD}^{(C'|(B'|(A'}T_{EFGH}^{E')|F')|G')J'}
\end{equation}
since again we can permute the symmetrized indices and use (\ref{eq:zz}).
The same procedure gives the following three identities with two
symmetrizations in the first, one symmetrization in the second, and no
symmetrization in the third expression

$$
    T_{J'K'ABCD}^{(B'|(A'}T_{EFGH}^{E')|F')J'K'}-T_{ABCD}^{J'K'(F'|(E'}
    T_{EFGHJ'K'}^{A')|B')}=0
$$

\begin{equation} \label{type10symmetry}
        T_{J'K'L'ABCD}^{(A'}T_{EFGH}^{E')J'K'L'}-
        T_{ABCD}^{J'K'L'(E'}T_{EFGHJ'K'L'}^{A')}=
        2T_{J'K'L'ABCD}^{(A'}T_{EFGH}^{E')J'K'L'}
\end{equation}

$$
        T_{J'K'L'M'ABCD}T_{EFGH}^{J'K'L'M'}-
        T_{ABCD}^{J'K'L'M'}T_{EFGHJ'K'L'M'}=0
$$

Next we look at symmetrizations of the lower indices. Due to the
above we only need to care about the cases where we have 1 or 3
symmetrizations in the upper indices. Thus we only need to look at
symmetrizations of the lower indices of (\ref{type30symmetry}) and
(\ref{type10symmetry}). Let us call a symmetrization of type
$\binom{n}{m}$ when we symmetrize in $n$ upper indices and $m$
lower indices. 
If $n+m$ is odd then, by permuting all the symmetrized pairs and by using
 (\ref{eq:zz}), also an odd number of times, on the contracted pairs, we
 see that such terms vanish. Hence only terms with $n+m$ even do not vanish
 and as $n=3$ or $n=1$,
 this implies that only terms of the types  $\binom{3}{3}$, $\binom{3}{1}$,
 $\binom{1}{3}$, and  $\binom{1}{1}$ can remain. These are

$$
\begin{array}{lll}\label{typesymmetry}
       & T_{JJ'(C|(B|(A}^{(C'|(B'|(A'}T_{E)|F)|G)}^{E')F')G')JJ'}\ \  , \quad T_{JKLJ'(A}^{(C'|(B'|(A'}T_{E)}^{E')|F')|G')JKLJ'}\ \ ,\\
       & T_{JJ'K'L'(C|(B|(A}^{(A'}T_{E)|F)|G)}^{E')JJ'K'L'}\   {\mbox{and}} \ \ \ 
        T_{JKLJ'K'L'(A}^{(A'}T_{E)}^{E')JKLJ'K'L'}
         \end{array} 
$$

The identity (\ref{eq:fun4}) holds if and only if all the above types of
symmetrizations vanish. Moreover noticing that the types
$\binom{1}{3}$ and $\binom{3}{1}$ are complex conjugates, we arrive
at the  lemma.

\end{proof}

The expressions obtained above seem nice but the problem is that
they cannot directly be converted into a tensorial expression in
any comfortable way.

\begin{lemma}\label{th:L3}
$$
T_{ABCD}^{A'B'C'D'}T_{EFGH}^{E'F'G'H'}=T_{ABCD}^{E'F'G'H'}T_{EFGH}^{A'B'C'D'}
$$
if and only if
\begin{equation}
\begin{array}{lll}\label{eq:fun42}
     &T_{jCB(A}^{C'B'(A'}T_{E)FG}^{E')F'G'j}-
    \frac{1}{4}\varepsilon_{BF}\bar\varepsilon^{B'F'}T_{jk(C|(A}^{(C'|(A'}T_{E)|G)}^{E')|G')jk}
    -\frac{1}{4}\varepsilon_{CG}\bar\varepsilon^{C'G'}T_{jk(B|(A}^{(B'|(A'}T_{E)|F)}^{E')|F')jk}\\\\
    &+\frac{1}{4}\varepsilon_{BC}\bar\varepsilon^{B'C'}T_{jk(F|(A}^{(F'|(A'}T_{E)|G)}^{E')|G')jk}
    -\frac{1}{4}\varepsilon_{CF}\bar\varepsilon^{C'F'}T_{jk(B|(A}^{(B'|(A'}T_{E)|G)}^{E')|G')jk}\\\\
    &+\frac{1}{4}\varepsilon_{FG}\bar\varepsilon^{F'G'}T_{jk(B|(A}^{(B'|(A'}T_{E)|C)}^{E')|C')jk}
    -\frac{1}{4}\bar\varepsilon^{B'G'}\varepsilon_{BG}T_{jk(C|(A}^{(C'|(A'}T_{E)|F)}^{E')|F')jk}
    \quad =0
\end{array}
\end{equation}
\end{lemma}

\begin{proof}
Notice that if we separate the type
$\binom{1}{1}$ term $T_{jCB(A}^{C'B'(A'}T_{E)FG}^{E')F'G'j}$ 
into symmetric and antisymmetric parts 4 times
according to (\ref{eq:sas}) in two pairs of primed indices and
then in two pairs of unprimed indices, we get an expression with
16 terms containing terms of the types $\binom{1}{1}$,
$\binom{1}{2}$, $\binom{2}{1}$, $\binom{3}{1}$, $\binom{1}{3}$,
$\binom{2}{2}$, $\binom{3}{2}$, $\binom{2}{3}$ and $\binom{3}{3}$.
Terms with an odd total number of contractions will vanish because of 
(\ref{eq:zz}). Therefore only terms of the types $\binom{1}{1}$, $\binom{3}{1}$, 
$\binom{1}{3}$, $\binom{2}{2}$, and $\binom{3}{3}$ remain.
Taking the symmetric/antisymmetric parts with respect to the pairs of
indices $BF$, $CG$, $B'F'$ and $C'G'$ gives

\begin{equation} \label{4sym-antisym-reduced}\begin{array}{lll}
    &T_{jCB(A}^{C'B'(A'}T_{E)FG}^{E')F'G'j}=T_{j(C|(B|(A}^{(C'|(B'|(A'}T_{E)|F)|G)}^{E')|F')|G')j}
    +\frac{1}{4}\varepsilon_{BF}\bar\varepsilon^{B'F'}T_{jKK'(C|(A}^{(C'|(A'}T_{E)|G)}^{E')|G')jKK'}\\\\
    &+\frac{1}{4}\varepsilon_{BF}\varepsilon_{CG}T_{jKL(A}^{(C'|(B'|(A'}T_{E)}^{E')|F')|G')jKL}
    +\frac{1}{4}\bar\varepsilon^{B'F'}\varepsilon_{CG}T_{jKK'(B|(A}^{(C'|(A'}T_{E)|F)}^{E')|G')jKK'}\\\\
    &+\frac{1}{4}\varepsilon_{BF}\bar\varepsilon^{C'G'}T_{jKK'(C|(A}^{(B'|(A'}T_{E)|G)}^{E')|F')jKK'}
    +\frac{1}{4}\bar\varepsilon^{B'F'}\bar\varepsilon^{C'G'}T_{jK'L'(C|(B|(A}^{(A'}T_{E)|F)|G)}^{E')jK'L'}\\\\
    &+\frac{1}{4}\varepsilon_{CG}\bar\varepsilon^{C'G'}T_{jKK'(B|(A}^{(B'|(A'}T_{E)|F)}^{E')|F')jKK'}
    +\frac{1}{16}\varepsilon_{BF}\bar\varepsilon^{B'F'}\varepsilon_{CG}\bar\varepsilon^{C'G'}
    T_{jKLK'L'(A}^{(A'}T_{E)}^{E')jKLK'L'}
\end{array}
\end{equation}

To rewrite the expression $\varepsilon_{BF}\bar\varepsilon^{C'G'}
T_{jk(C|(A}^{(B'|(A'}T_{E)|G)}^{E')|F')jk}+\varepsilon_{CG}\bar\varepsilon^{B'F'}
T_{jk(B|(A}^{(C'|(A'}T_{E)|F)}^{E')|G')jk}$ we apply (\ref{eq:sas2}) to the pair
$CF$ in the first term and to $G'B'$ in the second to write it
\begin{eqnarray}
&\varepsilon_{BC}\bar\varepsilon^{C'G'}T_{jk(F|(A}^{(B'|(A'}T_{E)|G)}^{E')|F')jk}+
\varepsilon_{CF}\varepsilon_{B}{}^{L}\bar\varepsilon^{C'G'}
T_{jk(L|(A}^{(B'|(A'}T_{E)|G)}^{E')|F')jk}
\nonumber \\
&+\varepsilon_{CG}\bar\varepsilon^{G'F'}T_{jk(B|(A}^{(C'|(A'}T_{E)|F)}^{E')|B')jk}+
\bar\varepsilon^{B'G'}\varepsilon_{CG}\bar\varepsilon_{L'}{}^{F'}
T_{jk(B|(A}^{(C'|(A'}T_{E)|F)}^{E')|L')jk}
\nonumber \\
=&\varepsilon_{BC}\bar\varepsilon^{C'G'}T_{jk(F|(A}^{(B'|(A'}T_{E)|G)}^{E')|F')jk}+
\varepsilon_{CF}\bar\varepsilon^{C'G'}
T_{jk(B|(A}^{(B'|(A'}T_{E)|G)}^{E')|F')jk}
\nonumber \\
&+\varepsilon_{CG}\bar\varepsilon^{G'F'}T_{jk(B|(A}^{(C'|(A'}T_{E)|F)}^{E')|B')jk}+
\bar\varepsilon^{B'G'}\varepsilon_{CG}
T_{jk(B|(A}^{(C'|(A'}T_{E)|F)}^{E')|F')jk} \nonumber
\end{eqnarray}

Next apply (\ref{eq:sas2}) on $G'B'$ in the first term, on $G'F'$ in the second, on
$FC$ in the third, and on $BC$ in the last. The expression then becomes
\begin{eqnarray}
&\varepsilon_{BC}\bar\varepsilon^{C'B'}T_{jk(F|(A}^{(G'|(A'}T_{E)|G)}^{E')|F')jk}+
\bar\varepsilon^{G'B'}\varepsilon_{BC}\bar\varepsilon^{C'}{}_{L'}
T_{jk(F|(A}^{(L'|(A'}T_{E)|G)}^{E')|F')jk}
\nonumber \\
&+\varepsilon_{CF}\bar\varepsilon^{C'F'}T_{jk(B|(A}^{(B'|(A'}T_{E)|G)}^{E')|G')jk}
+\bar\varepsilon^{G'F'}\varepsilon_{CF}\bar\varepsilon^{C'}{}_{L'}
T_{jk(B|(A}^{(B'|(A'}T_{E)|G)}^{E')|L')jk}
\nonumber \\
&+\varepsilon_{FG}\bar\varepsilon^{G'F'}T_{jk(B|(A}^{(C'|(A'}T_{E)|C)}^{E')|B')jk}+
\varepsilon_{FC}\varepsilon^{L}{}_{G}\bar\varepsilon^{G'F'}
T_{jk(B|(A}^{(C'|(A'}T_{E)|L)}^{E')|B')jk}
\nonumber \\
&+\bar\varepsilon^{B'G'}\varepsilon_{BG}T_{jk(C|(A}^{(C'|(A'}T_{E)|F)}^{E')|F')jk}+
\varepsilon_{BC}\bar\varepsilon^{B'G'}\varepsilon^{L}{}_{G}
T_{jk(L|(A}^{(C'|(A'}T_{E)|F)}^{E')|F')jk}
\nonumber \\
=&\varepsilon_{BC}\bar\varepsilon^{C'B'}T_{jk(F|(A}^{(G'|(A'}T_{E)|G)}^{E')|F')jk}-
\bar\varepsilon^{G'B'}\varepsilon_{BC}
T_{jk(F|(A}^{(C'|(A'}T_{E)|G)}^{E')|F')jk}
\nonumber \\
&+\varepsilon_{CF}\bar\varepsilon^{C'F'}T_{jk(B|(A}^{(B'|(A'}T_{E)|G)}^{E')|G')jk}-
\bar\varepsilon^{G'F'}\varepsilon_{CF}
T_{jk(B|(A}^{(B'|(A'}T_{E)|G)}^{E')|C')jk}
\nonumber \\
&+\varepsilon_{FG}\bar\varepsilon^{G'F'}T_{jk(B|(A}^{(C'|(A'}T_{E)|C)}^{E')|B')jk}-
\varepsilon_{FC}\bar\varepsilon^{G'F'}
T_{jk(B|(A}^{(C'|(A'}T_{E)|G)}^{E')|B')jk}
\nonumber \\
&+\bar\varepsilon^{B'G'}\varepsilon_{BG}T_{jk(C|(A}^{(C'|(A'}T_{E)|F)}^{E')|F')jk}-
\varepsilon_{BC}\bar\varepsilon^{B'G'}
T_{jk(G|(A}^{(C'|(A'}T_{E)|F)}^{E')|F')jk} \nonumber
\end{eqnarray}

In the last expression terms 2 and 8 cancel as do terms 4 and 6. Hence
\begin{eqnarray}\label{eq:om2}
&\varepsilon_{BF}\bar\varepsilon^{C'G'}
T_{jk(C|(A}^{(B'|(A'}T_{E)|G)}^{E')|F')jk}+\varepsilon_{CG}\bar\varepsilon^{B'F'}
T_{jk(B|(A}^{(C'|(A'}T_{E)|F)}^{E')|G')jk}
\nonumber \\
=&\varepsilon_{BC}\bar\varepsilon^{C'B'}T_{jk(F|(A}^{(G'|(A'}T_{E)|G)}^{E')|F')jk}
+\varepsilon_{CF}\bar\varepsilon^{C'F'}T_{jk(B|(A}^{(B'|(A'}T_{E)|G)}^{E')|G')jk}
\nonumber \\
&+\varepsilon_{FG}\bar\varepsilon^{G'F'}T_{jk(B|(A}^{(C'|(A'}T_{E)|C)}^{E')|B')jk}
+\bar\varepsilon^{B'G'}\varepsilon_{BG}T_{jk(C|(A}^{(C'|(A'}T_{E)|F)}^{E')|F')jk} 
\end{eqnarray}

This together with (\ref{4sym-antisym-reduced}) gives
$$
\begin{array}{lll}
    &T_{jBC(A}^{B'C'(A'}T_{E)FG}^{E')F'G'j}-
    \frac{1}{4}\varepsilon_{BF}\bar\varepsilon^{B'F'}T_{jk(C|(A}^{(C'|(A'}T_{E)|G)}^{E')|G')jk}
    -\frac{1}{4}\varepsilon_{CG}\bar\varepsilon^{C'G'}T_{jk(B|(A}^{(B'|(A'}T_{E)|F)}^{E')|F')jk}\\\\
   & - \frac{1}{4}\varepsilon_{BC}\bar\varepsilon^{C'B'}T_{jk(F|(A}^{(G'|(A'}T_{E)|G)}^{E')|F')jk}
    -\frac{1}{4}\varepsilon_{CF}\bar\varepsilon^{C'F'}T_{jk(B|(A}^{(B'|(A'}T_{E)|G)}^{E')|G')jk}\\\\
    &-\frac{1}{4}\varepsilon_{FG}\bar\varepsilon^{G'F'}T_{jk(B|(A}^{(C'|(A'}T_{E)|C)}^{E')|B')jk}
    -\frac{1}{4}\bar\varepsilon^{B'G'}\varepsilon_{BG}T_{jk(C|(A}^{(C'|(A'}T_{E)|F)}^{E')|F')jk}\\\\
    =&    T_{j(C|(B|(A}^{(C'|(B'|(A'}T_{E)|F)|G)}^{E')|F')|G')j}
    +\frac{1}{4}\varepsilon_{BF}\varepsilon_{CG}T_{jKL(A}^{(C'(B'|(A'}T_{E)}^{E')|F')|G')jKL}\\\\
    &+\frac{1}{4}\bar\varepsilon^{B'F'}\bar\varepsilon^{C'G'}T_{jK'L'(C|(B|(A}^{(A'}T_{E)|F)|G)}^{E')jK'L'}
        +\frac{1}{16}\varepsilon_{BF}\bar\varepsilon^{B'F'}\varepsilon_{CG}\bar\varepsilon^{C'G'}
    T_{jkl(A}^{(A'}T_{E)}^{E')jkl} \nonumber
\end{array}
$$

As an expression vanishes if and only if all its symmetric/antisymmetric parts vanish,
application of Lemma \ref{th:L2} together with $\varepsilon_{AB}=-\varepsilon_{BA}$ 
completes the proof.
\end{proof}

We still have terms of type $\binom{2}{2}$ but we can eliminate them:

\begin{lemma}\label{th:L4}
$$
T_{ABCD}^{A'B'C'D'}T_{EFGH}^{E'F'G'H'}=T_{ABCD}^{E'F'G'H'}T_{EFGH}^{A'B'C'D'}
$$
if and only if
\begin{equation}
\begin{array}{lll}\label{eq:fun43}
     &T_{jbc(A}^{(A'}T_{E)fg}^{E')j}-
    \frac{1}{4}g_{bf}T_{jkc(A}^{(A'}T_{E)g}^{E')jk}
    -\frac{1}{4}g_{cg}T_{jkb(A}^{(A'}T_{E)f}^{E')jk}
    - \frac{1}{4}g_{bg}T_{jkc(A}^{(A'}T_{E)f}^{E')jk}
    -\frac{1}{4}g_{cf}T_{jkb(A}^{(A'}T_{E)g}^{E')jk}\\\\
    &+\frac{1}{4}g_{bc}T_{jkf(A}^{(A'}T_{E)g}^{E')jk}
    +\frac{1}{4}g_{fg}T_{jkb(A}^{(A'}T_{E)c}^{E')jk}
    +\frac{1}{8}(g_{bf}g_{cg}+g_{bg}g_{cf}-g_{bc}g_{fg})T_{jkl(A}^{(A'}T_{E)}^{E')jkl}
    \quad =0
\end{array}
\end{equation}
\end{lemma}

\begin{proof}
Consider the expression $T_{jkB(A}^{B'(A'}T_{E)F}^{E')F'jk}$ and split into symmetric 
and antisymmetric parts with respect to the index pairs $B'F'$ and $BF$. This gives
$$
\begin{array}{lll}
T_{jkB(A}^{B'(A'}T_{E)F}^{E')F'jk}=&T_{jk(B|(A}^{(B'|(A'}T_{E)|F)}^{E')|F')jk}+\frac12\varepsilon_{BF}
T_{jkL(A}^{(B'|(A'}T_{E)}^{E')|F')jkL}\\\\
    &+\frac12\bar\varepsilon^{B'F'}T_{jkL'(B|(A}^{(A'}T_{E)|F)}^{E')jkL'}+
    \frac{1}{4}\varepsilon_{BF}\bar\varepsilon^{B'F'}T_{jkl(A}^{(A'}T_{E)}^{E')jkl} 
\end{array}
$$

Here in the terms with coefficient $\frac12$ we have an odd number
of contractions, so these terms vanish and we are left with

\begin{equation}\label{eq:s21}
    T_{jk(B|(A}^{(B'|(A'}T_{E)|F)}^{E')|F')jk}=T_{jkB(A}^{B'(A'}T_{E)F}^{E')F'jk}-\frac{1}{4}\epsilon_{BF}\epsilon^{B'F'}T_{jkl(A}^{(A'}T_{E)}^{E')jkl}
\end{equation}

Doing the same for all type $\binom{2}{2}$ terms in  (\ref{eq:fun42}) gives terms with
$T_{jkl(A}^{(A'}T_{E)}^{E')jkl}$ multiplied by
$\frac{1}{16}\varepsilon_{BF}\bar\varepsilon^{B'F'}\varepsilon_{CG}\bar\varepsilon^{C'G'}\ $,
$    \frac{1}{16}\varepsilon_{CG}\bar\varepsilon^{C'G'}\varepsilon_{BF}\bar\varepsilon^{B'F'}\ $,
 $   -\frac{1}{16}\varepsilon_{BC}\bar\varepsilon^{B'C'}\varepsilon_{FG}\bar\varepsilon^{F'G'}\ $,
   $ \frac{1}{16}\varepsilon_{CF}\bar\varepsilon^{C'F'}\bar\varepsilon^{B'G'}\varepsilon_{BG}\ $,
   $ -\frac{1}{16}\varepsilon_{FG}\bar\varepsilon^{F'G'}\varepsilon_{BC}\bar\varepsilon^{B'C'}$ and
   $ \frac{1}{16}\bar\varepsilon^{B'G'}\varepsilon_{BG}\varepsilon_{CF}\bar\varepsilon^{C'F'}$
 respectively.
Substituting all this into Lemma \ref{th:L3} and lowering the indices $B'$, $C'$, $F'$ and
$G'$ completes the proof.

\end{proof}

\begin{lemma}\label{th:L5}
$$
T_{ABCD}^{A'B'C'D'}T_{EFGH}^{E'F'G'H'}=T_{ABCD}^{E'F'G'H'}T_{EFGH}^{A'B'C'D'}
$$
if and only if (\ref{eq:RaiS}) is satisfied.
\end{lemma}

\begin{proof}
We apply (\ref{eq:ds}) to the expression (\ref{eq:fun43}) to get rid of all spinor indices.
This gives, after lowering $E$ and $E'$

\begin{equation}
\begin{array}{lll}\label{eq:fun44}
     &T_{jbc(a}T_{e)fg}{}^{j}-
    \frac{1}{4}g_{bf}T_{jkc(a}T_{e)g}{}^{jk}
    -\frac{1}{4}g_{cg}T_{jkb(a}T_{e)f}{}^{jk}
    - \frac{1}{4}g_{bg}T_{jkc(a}T_{e)f}{}^{jk}
    -\frac{1}{4}g_{cf}T_{jkb(a}T_{e)g}{}^{jk}\\
    &+\frac{1}{4}g_{bc}T_{jkf(a}T_{e)g}{}^{jk}
    +\frac{1}{4}g_{fg}T_{jkb(a}T_{e)c}{}^{jk}
    +\frac{1}{8}(g_{bf}g_{cg}+g_{bg}g_{cf}-g_{bc}g_{fg})T_{jkla}T_{e}{}^{jkl}\\
    &-\frac{1}{4}g_{ae}T_{jkbc}T_{fg}{}^{jk}
    -\frac{1}{16}g_{ae}(-g_{bf}T_{jklc}T_{g}{}^{jkl}-g_{cg}T_{jklb}T_{f}{}^{jkl}
    -g_{bg}T_{jklc}T_{f}{}^{jkl}-g_{cf}T_{jklb}T_{g}{}^{jkl}\\
    &+g_{bc}T_{jklf}T_{g}{}^{jkl}+g_{fg}T_{jklb}T_{c}{}^{jkl}
    +\frac{1}{2}(g_{bf}g_{cg}+g_{bg}g_{cf}-g_{bc}g_{fg})T_{jklm}T^{jklm})
    \quad =0
\end{array}
\end{equation}
which is equivalent to (\ref{eq:fun43}). We simplify by adding some trace. Taking the trace over $c$ and $g$ we get
$$
\begin{array}{lll}\label{eq:fun44tr}
     &T_{jkb(a}T_{e)f}{}^{jk}-
    \frac{1}{4}g_{bf}T_{jkla}T_{e}{}^{jkl}
    -T_{jkb(a}T_{e)f}{}^{jk}
    - \frac{1}{4}T_{jkb(a}T_{e)f}{}^{jk}
    -\frac{1}{4}T_{jkb(a}T_{e)f}{}^{jk}\\
    &+\frac{1}{4}T_{jkf(a}T_{e)b}{}^{jk}
    +\frac{1}{4}T_{jkb(a}T_{e)f}{}^{jk}
    +\frac{1}{8}(4g_{bf}+g_{bf}-g_{bf})T_{jkla}T_{e}{}^{jkl}
    -\frac{1}{4}g_{ae}T_{jklb}T_{f}{}^{jkl}\\
    &-\frac{1}{16}g_{ae}(-g_{bf}T_{jklm}T^{jklm}-4T_{jklb}T_{f}{}^{jkl}
    -T_{jklb}T_{f}{}^{jkl}-T_{jklb}T_{f}{}^{jkl}\\
    &+T_{jklf}T_{b}{}^{jkl}+T_{jklb}T_{f}{}^{jkl}
    +\frac{1}{2}(4g_{bf}+g_{bf}-g_{bf})T_{jklm}T^{jklm})\\
    =&\frac{1}{4}g_{bf}(T_{jkla}T_{e}{}^{jkl}-\frac{1}{4}g_{ae}T_{jklm}T^{jklm})
    \quad =0 
\end{array}
$$
Therefore
\begin{equation}\label{eq:fun44tr2}
T_{jkla}T_{e}{}^{jkl}-\frac{1}{4}g_{ae}T_{jklm}T^{jklm} =0
\end{equation}
Substituting this into all terms (\ref{eq:fun44}) containing 3 traces
we get
\begin{equation}
\begin{array}{lll}\label{eq:fun45}
     &T_{jbc(a}T_{e)fg}{}^{j}-
    \frac{1}{4}g_{bf}T_{jkc(a}T_{e)g}{}^{jk}
    -\frac{1}{4}g_{cg}T_{jkb(a}T_{e)f}{}^{jk}
    - \frac{1}{4}g_{bg}T_{jkc(a}T_{e)f}{}^{jk}\\
    &-\frac{1}{4}g_{cf}T_{jkb(a}T_{e)g}{}^{jk}
    +\frac{1}{4}g_{bc}T_{jkf(a}T_{e)g}{}^{jk}
    +\frac{1}{4}g_{fg}T_{jkb(a}T_{e)c}{}^{jk}\\
    &-\frac{1}{4}g_{ae}(T_{jkbc}T_{fg}{}^{jk}
    -\frac{1}{8}(g_{bf}g_{cg}+g_{bg}g_{cf}-g_{bc}g_{fg})T_{jklm}T^{jklm}))
    \quad =0
\end{array}
\end{equation}
That this is equivalent to (\ref{eq:fun44}) is guaranteed by the fact that we also from
this equation can obtain (\ref{eq:fun44tr2}), e.g. by taking traces over $bf$ and $cg$.
Therefore (\ref{eq:fun45}) is equivalent to (\ref{eq:fun43}).

\end{proof}

\begin{lemma}\label{th:L6} If $\quad 
T_{ABCD}^{A'B'C'D'}T_{EFGH}^{E'F'G'H'}=T_{ABCD}^{E'F'G'H'}T_{EFGH}^{A'B'C'D'}\quad$
then 
\begin{eqnarray}\label{eq:as4}
&T_{jbc[a}T_{e]fg}{}^{j}={1\over 4}g_{bf}T_{jkc[a}T_{e]g}{}^{jk}+
{1\over 4}g_{cg}T_{jkb[a}T_{e]f}{}^{jk} 
+g_{[f|[a}T_{e]jk(c}T_{g)|b]}{}^{jk}\nonumber \\
&+g_{[g|[a}T_{e]jk(b}T_{f)|c]}{}^{jk}
+{1\over 16}(g_{bf}g_{c[a}g_{e]g}+ g_{cg}g_{b[a}g_{e]f})
T_{jklm}T^{jklm}
\end{eqnarray}
\end{lemma}

\begin{proof}
Expressing the anti-symmetric part in spinors we have
$$
T_{jbc[a}T_{e]fg}{}^{j}={1\over 2}\bar\varepsilon_{A'E'}T_{jK'bc(A}T_{E)fg}{}^{jK'}+
{1\over 2}\varepsilon_{AE}T_{jKbc(A'}T_{E')fg}{}^{jK}
$$
Taking symmetric and anti-symmetric parts we get terms of types $\binom{n}{m}$
with $1\le n+m \le 5$, $n\le 3$ and $m\le 3$. Terms with $n+m$ odd vanish identically while
terms of types $\binom{3}{1}$, $\binom{1}{3}$ and $\binom{1}{1}$ vanish by lemma
\ref{th:L2}. Hence only types $\binom{2}{2}$, $\binom{2}{0}$ and $\binom{0}{2}$ remain
and this gives
\begin{equation}
\begin{array}{lll}\label{eq:fun46}
     &T_{jbc[a}T_{e]fg}{}^{j}=
    {1\over 4}\bar\varepsilon_{A'E'}\varepsilon_{CG}T_{jk(C'|(A|(B'|(B}T_{F)|F')|E)|G')}{}^{jk}\\
    &+{1\over 4}\bar\varepsilon_{A'E'}\varepsilon_{BF}T_{jk(B'|(A|(C'|(C}T_{G)|G')|E)|F')}{}^{jk}
   + {1\over 16}\bar\varepsilon_{A'E'}\varepsilon_{CG}\bar\varepsilon_{B'F'}\bar\varepsilon_{C'G'}
    T_{jkL'M'(B|(A}T_{E)|F)}{}^{jkL'M'}\\
   &+{1\over 16}\bar\varepsilon_{A'E'}\varepsilon_{BF}\bar\varepsilon_{B'F'}\bar\varepsilon_{C'G'}
    T_{jkL'M'(C|(A}T_{E)|G)}{}^{jkL'M'} + CC
    \end{array}
\end{equation}
where $CC$ means complex conjugate. Next observe that by (\ref{eq:sas})
\begin{equation}
\begin{array}{lll}\label{eq:fun47}
     &T_{jkL'b(A}T_{E)f}{}^{jkL'}=T_{jkL'(B'|(B|(A}T_{E)|F)|F')}{}^{jkL'}
     +{1\over 2}\varepsilon_{BF}T_{jkl(B'|(A}T_{E)|F')}{}^{jkl}\\
    &+{1\over 2}\bar\varepsilon_{B'F'}T_{jkL'M'(B|(A}T_{E)|F)}{}^{jkL'M'}
    +{1\over 4}\varepsilon_{BF}\bar\varepsilon_{B'F'}T_{jklM'(A}T_{E)}{}^{jklM'}\\
    &={1\over 2}\bar\varepsilon_{B'F'}T_{jkL'M'(B|(A}T_{E)|F)}{}^{jkL'M'}
    \end{array}
\end{equation}
where in the last step the first and the last terms vanish identically and the second term
vanishes by lemma \ref{th:L2}. Furthermore
\begin{equation}\label{eq:fun48}
   \bar\varepsilon_{A'E'}T_{jkL'b(A}T_{E)f}{}^{jkL'} + CC=
2T_{jkb[a}T_{e]f}{}^{jk}  
\end{equation}
On the other hand, by (\ref{eq:om2}), (\ref{eq:s21}) and lemma \ref{th:L2} we have
\begin{equation}
\begin{array}{lll}\label{eq:fun49}
     &\bar\varepsilon_{A'E'}\varepsilon_{CG}T_{jk(C'|(A|(B'|(B}T_{F)|F')|E)|G')}{}^{jk} + CC\\
     =&-g_{ac}T_{jke(B'|(B}T_{F)|F')g}{}^{jk} + g_{ag}T_{jkc(B'|(B}T_{F)|F')e}{}^{jk}\\
     &-g_{eg}T_{jka(B'|(B}T_{F)|F')c}{}^{jk} + g_{ce}T_{jka(B'|(B}T_{F)|F')g}{}^{jk}
    \end{array}
\end{equation}
Applying (\ref{eq:ds}) to (\ref{eq:fun49}) and substituting the result together with
(\ref{eq:fun47}) and (\ref{eq:fun48}) into (\ref{eq:fun46}) we obtain the formula 
(\ref{eq:as4}).

\end{proof}

{\it Proof of Theorem \ref{th:RaiBR2}}

We have
$$
T_{jabc}T_{efg}{}^{j}=T_{jbc(a}T_{e)fg}{}^{j}+T_{jbc[a}T_{e]fg}{}^{j}
$$
where $T_{jbc(a}T_{e)fg}{}^{j}$ is given by (\ref{eq:RaiS}) and $T_{jbc[a}T_{e]fg}{}^{j}$ 
by (\ref{eq:as4}). Adding these expressions it is not obvious that (\ref{eq:Rai}) is
obtained but since the expression must be symmetric in $abc$ and in $efg$ it equals
its symmetric part with the respect to $abc$ and $efg$. Writing out the full expression and 
(with $efg$ raised) taking such symmetric parts of each term, only terms of types 
$g_{(a}{}^{(e}T_{bc)jk}T^{fg)jk}$, $g_{(a}{}^{(e}T_{|jk|b}{}^{f}T_{c)}{}^{g)jk}$,
$g_{(ab}T_{c)jk}{}^{(e}T^{fg)jk}$, $g^{(ef}T_{jk(ab}T_{c)}{}^{g)jk}$,
$g_{(ab}g_{c)}{}^{(e}g^{fg)}T_{jklm}T^{jklm}$ and
$g_{(a}{}^{(e}g_b{}^f g_{c)}{}^{g)}T_{jklm}T^{jklm}$ can occur. Simply counting
the coefficients gives the formula (\ref{eq:Rai}).

Note that by lemmas \ref{th:L5} and \ref{th:L6}, (\ref{eq:Rai}) is obviously implied by
(\ref{eq:fun4}) but the converse is also true since  (\ref{eq:Rai})  implies
 (\ref{eq:RaiS})  (by taking a symmetric part) and since  (\ref{eq:RaiS}) implies
 (\ref{eq:fun4}) by lemma \ref{th:L5}. Hence, by lemma \ref{th:L1}, the theorem is
 proved. Note that this also proves Theorem \ref{th:RaiBR}.

{\hfil\hfill$\square$\endtrivlist}

\section{Discussion}

We have presented the first Rainich type result for higher rank superenergy tensors.
It seems clear that it is the complexity of the derivation that has prevented it from
being found before. Still, the identity is only quadratic and on the form (\ref{eq:Rai})  one
sees clearly all the expected symmetries. We believe spinor methods are probably much 
easier to use than tensor methods. If a tensorial proof of our result can be found, then one may 
consider generalizations to arbitrary dimension or arbitrary signature of the metric. 
Various generalizations of the contracted identity (\ref{eq:BR3c}) were given in \cite{EW}. 
There are many other possible generalizations. With spinor methods one can study more
general superenergy tensors in the 4-dimensional Lorenzian case and look for necessary
and sufficient identities. It would be interesting to see if general causal tensors can always
be expressed in terms of supernergy tensors as in the rank-2 case \cite{BS}.
 From our results one may also try to find necessary and
sufficient identities for the different Petrov types of the Weyl tensor.
Finally, as mentioned in the introduction, results on differential conditions for higher
rank superenergy tensors will be presented in a forthcoming paper.

\subsection*{Acknowledgements}
We thank Jos\'e Senovilla for many useful suggestions and comments and Jonas Bergman
and Brian Edgar  for discussions on spinor and tensor identities.


\begin{thebibliography}{99}
    
\bibitem{B}
    Bergqvist G 1999
    Positivity of general superenergy tensors
    \textit{Commun. Math. Phys.}
    \textbf{207} 467--479

\bibitem{BH}
    Bergqvist G and H{\"o}glund A  2002
    Algebraic Rainich theory and antisymmetrization in higher dimensions
    \textit{Class. Quantum Grav.}
    \textbf{19} 3341--3355

\bibitem{BS}
    Bergqvist G and Senovilla J  M  M  2001
    Null cone preserving maps, causal tensors and algebraic Rainich theory
    \textit{Class. Quantum Grav.}
    \textbf{18} 5299--5325

\bibitem{EW}
    Edgar S B and Wingbrandt O  2003
    Old and new results for superenergy tensors from dimensionally dependent tensor identities
    \textit{J. Math. Phys.}
    \textbf{44} 6140--6159

\bibitem{MW}
    Misner C W and Wheeler J A 1957
    Classical physics as geometry
    \textit{Ann. Phys., NY}
    \textbf{2} 525--603

\bibitem{PR} 
	Penrose R and Rindler W 1984
{\it Spinors and spacetime} vol 1 Cambridge Univ. Press

\bibitem{R}
    Rainich G Y 1925
    Electrodynamics in the general relativity theory
    \textit{Trans. Am. Math. Soc.}
    \textbf{27} 106--136
  
\bibitem{S}
    Senovilla J M M 2000
    Super-energy tensors
    \textit{Class. Quantum Grav.}
    \textbf{17} 2799--2841

\end{thebibliography}
\end{document}